\documentclass[aps,prl,reprint,showpacs,superscriptaddress]{revtex4-1}
\usepackage{graphicx}
\usepackage{dcolumn}
\usepackage{bm}
\usepackage{amssymb}
\usepackage{amsmath}
\usepackage{subfigure}
\usepackage{epstopdf}
\usepackage{float}
\usepackage{mathrsfs}
\begin{document}
\title{Magnetostatics of Magnetic Skyrmion Crystals}
\author{Ren Qin}
\affiliation{School of Physics, Nankai University, Tianjin 300071, China}
\author{Yong Wang}
\email[]{yongwang@nankai.edu.cn}
\affiliation{School of Physics, Nankai University, Tianjin 300071, China}

\begin{abstract}
Magnetic skyrmion crystals are topological magnetic textures arising in the chiral ferromagnetic materials with Dzyaloshinskii-Moriya interaction. The magnetostatic fields generated by magnetic skyrmion crystals are first studied by micromagnetic simulations. For N\'eel-type skyrmion crystals, the fields will vanish on one side of the crystal plane, which depend on the helicity; while for Bloch-type skyrmion crystals, the fields will distribute over both sides, and are identical for the two helicities. These features and the symmetry relations of the magetostatic fields are understood from the magnetic scalar potential and magnetic vector potential of the hybridized triple-$\mathbf{Q}$ state. The possibility to construct magnetostatic field at nanoscale by stacking chiral ferromagnetic layers with magnetic skyrmion crystals is also discussed, which may have potential applications to trap and manipulate neutral atoms with magnetic moments. 
\end{abstract}
\maketitle
The lack of spatial inversion symmetry in chiral ferromagnets can give rise to the anistropic exchange interaction between the neighbouring magnetic moments, \emph{i.e.} the Dzyaloshinskii-Moriya (DM) mechanism\cite{Dzya,Moriya}. In contrast to the Heisenberg exchange interaction, which stabilizes the collinear magnetic structure with minimized free energy, DM interaction prefers non-collinear magnetic structures and enables the chiral ferromagnets to host the topological-protected magnetic skyrmions\cite{PRL2001,Nature2006,Science2009,Nature2010,NatMat2010,NatPhys2011}.  
Magnetic skyrmions are particle-like topological defects in the magnetization configuration, and their swirling structures are characterized by topological skyrmion numbers\cite{NatNano2013}. The exchange coupling between the magnetic skyrmions and the conduction electrons can further result in the exotic dynamics of emergent electromagnetic field\cite{NatNano2013,NatPhys2012}, such as topological Hall effect\cite{THE1,THE2,THE3,THE4} and skyrmion Hall effect\cite{SkHE1,SkHE2}. The attractive properties of magnetic skyrmions have been intensivley utilized to design and develop skyrmion-based topological electronics devices\cite{NRM1,NRM2,Proc}.

The magnetostatic field distribution generated by magnetic skyrmions, as governed by the Maxwell equations, is one of the fundamental physical features of these topological objects. Indeed, one important way to observe the magnetic skyrmions or other magnetised microstructures of magnetization is to detect their magnetic field profiles with various sensing techniques, including Lorentz transmission electron microscopy\cite{Nature2010}, magnetic force microscopy\cite{NC2017,arX2017}, nitrogen-vacancy magnetometry\cite{NC2013,arX2016,RMP2017}, \emph{etc}. Understanding the magnetostatic features of magnetic skyrmions will also be meaningful for designing skyrmion-based electronics devices\cite{NRM1,NRM2,Proc}. Furthermore, magnetic skyrmions have the potential applications to design magnetic microtraps, which are used to trap and manipulate ultracold atoms\cite{JPD1999,RMP2007}. In this Letter, we will investigate the magnetostatic fields generated by magnetic skyrmion crystals (SkXs) with different helicities, and show the possibility to construct the field distributions at nanoscale through stacking the chiral ferromagnet films for further applications. 

\begin{figure*}
\includegraphics[scale=0.62]{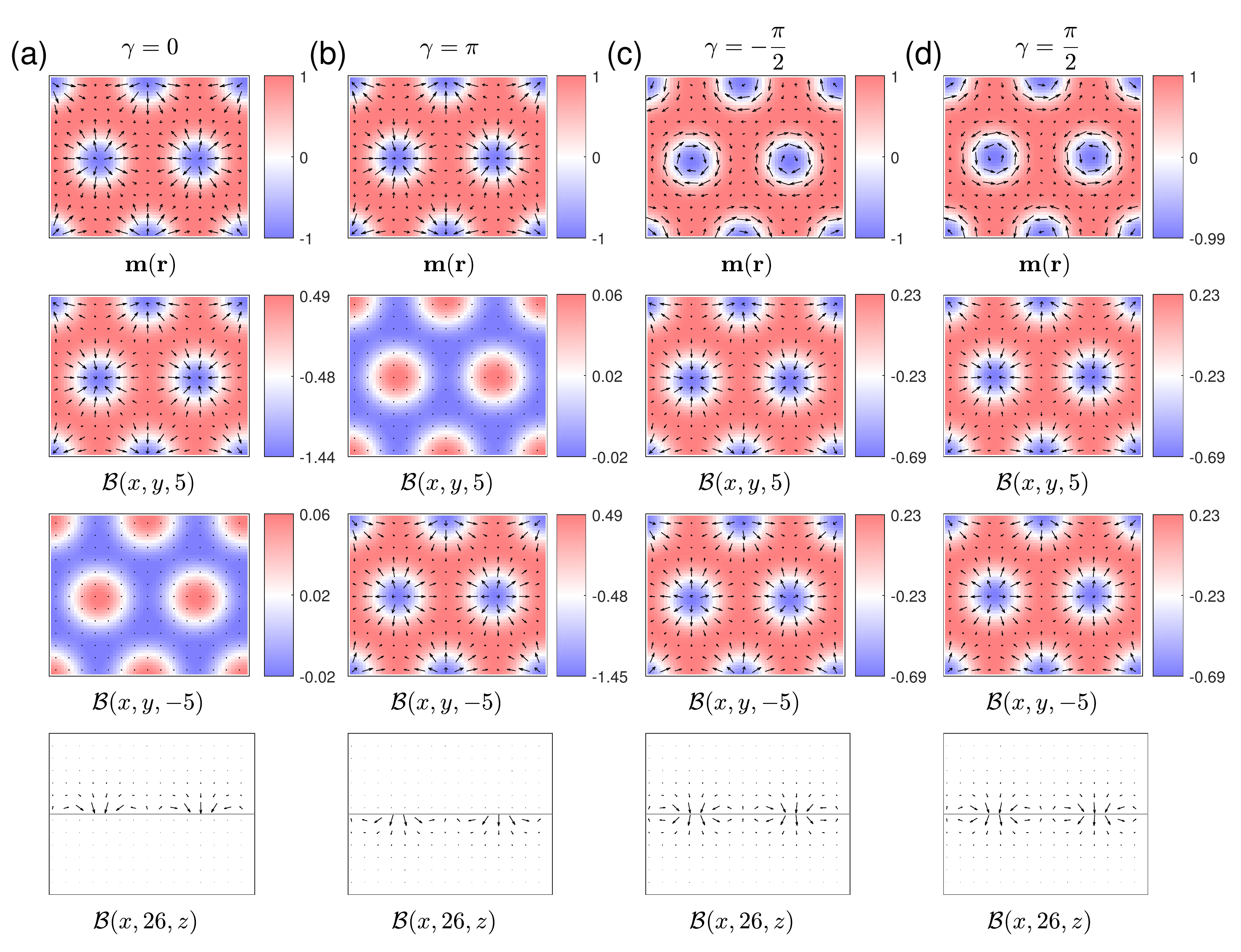}
\caption{(Color online) The magnetization configurations $\mathbf{m}(\mathbf{r})$ and the associated magnetostatic field distributions $\bm{\mathcal{B}}(\mathbf{r})$ at the $x$-$y$ planes at $z=\pm 5$ and the $x$-$z$ plane at $y=26$ for magnetic skyrmion crystals with different helicities $\gamma$. (a) $\gamma=0$; (b) $\gamma=\pi$; (c) $\gamma=-\frac{\pi}{2}$; (d) $\gamma=\frac{\pi}{2}$. The lattice constant is set as $1$ here.}\label{Fig1}
\end{figure*}

We consider a two-dimensional chiral ferromagnetic film placed in the external magnetic field, which can host magnetic skyrmion crystals\cite{NatNano2013,NatNano2013-2}. Its energy functional in terms of the normalized magnetic moments $\{\mathbf{m}_{i}\}$ on the discretized square lattice is given as
\begin{eqnarray}
\mathscr{E}[\{\mathbf{m}_{i}\}]=-J\sum_{\langle i,j\rangle}\mathbf{m}_{i}\cdot\mathbf{m}_{j}-\mathbf{D}\cdot\sum_{\langle i,j\rangle}\mathbf{m}_{i}\times\mathbf{m}_{j}-\mathbf{B}\cdot\sum_{i}\mathbf{m}_{i}.\nonumber\\\label{SkyMod}
\end{eqnarray}
Here, $\mathbf{m}_{i}$ denotes the normalized magnetic moment at lattice site $i$, and the summation $\langle i,j\rangle$ is over the nearest lattice sites $i$ and $j$; the first term in (\ref{SkyMod}) describes the ferromagnetic exchange interaction, where $J>0$ is the interaction strength; the second term in (\ref{SkyMod}) describes the DM interaction, where the form of $\mathbf{D}$ can be either $D\hat{\mathbf{r}}_{ij}$ or $D\hat{\mathbf{r}}_{ij}\times\hat{\mathbf{e}}_{z}$, with the notations $\mathbf{r}_{ij}=\mathbf{r}_{j}-\mathbf{r}_{i}$ and $\hat{\mathbf{e}}_{z}=(0,0,1)$; the third term in (\ref{SkyMod}) describes the Zeeman effect, where $\mathbf{B}$ is the external magnetic field. 

For a given parameter set $\{J,\mathbf{D},\mathbf{B}\}$, the stable magnetization configuration $\{\mathbf{m}_{i}^{0}\}$ is achieved by minimizing the energy functional $\mathscr{E}$ \emph{via} Landau-Lifshitz-Gilber(LLG) equation. Depending on the relative direction of $\mathbf{D}$ and $\mathbf{r}_{i,j}$, the obtained SkXs can be classified with four different helicities\cite{NatNano2013} ($\gamma=0,\pi$ for N\'eel-type SkXs and $\gamma=\pm\frac{\pi}{2}$ for Bloch-type SkXs), where $\hat{\mathbf{D}}\cdot\hat{\mathbf{r}}_{i,j}=\sin\gamma$ and $\hat{\mathbf{D}}\cdot(\hat{\mathbf{r}}_{i,j}\times\hat{\mathbf{e}}_{z})=\cos\gamma$. The resulting dimensionless magnetic field $\bm{\mathcal{B}}(\mathbf{r})$ is the summation over the magnetic dipole field generated by each magnetic moment $\mathbf{m}_{i}^{0}$, \emph{i.e.}
\begin{eqnarray}
\bm{\mathcal{B}}(\mathbf{r})=\sum_{i}\frac{3(\mathbf{m}_{i}^{0}\cdot\hat{\bm{\mathcal{R}}}_{i})\hat{\bm{\mathcal{R}}}_{i}-\mathbf{m}_{i}^{0}}{\mathcal{R}_{i}^{3}}.\label{StrayField}
\end{eqnarray}  
Here, $\bm{\mathcal{R}}_{i}=\mathbf{r}-\mathbf{r}_{i}$ denotes the displacement vector from the $i$th lattice site $\mathbf{r}_{i}$ to the spatial point $\mathbf{r}$. 

The magnetic skyrmion crystals with four different helicities $\gamma$ on a $60\times 52$ square lattice and the associated magnetic field distributions are obtained numerically and shown in Fig.~\ref{Fig1}. Here, we set\cite{NatNano2013-2} $J=1$~meV, $D=\pm0.238$~meV, $\mathbf{B}=(0,0,0.035)$~meV, and the periodic boundary condition is exploited. The magnetic moment in the center of each skyrmion will point towards the $-\hat{\mathbf{e}}_{z}$ direction, which is opposite to the applied magnetic field. As expected, all the four calculated magnetostatic fields have the same period as the original SkXs, and the field strength will decay at distance away from the crystal plane at $z=0$. Impressively, for the N\'eel-type SkXs with helicity $\gamma=0$ ($\gamma=\pi$), the field strength in the upper half-space $z>0$ is much stronger(weaker) than that in the lower half-space $z<0$, and the field components satisfy the symmetry relations $\mathcal{B}_{\gamma=0,x/y}(x,y,z)=-\mathcal{B}_{\gamma=\pi,x/y}(x,y,-z)$ and $\mathcal{B}_{\gamma=0,z}(x,y,z)=\mathcal{B}_{\gamma=\pi,z}(x,y,-z)$, as shown in Fig.~\ref{Fig1}(a)(b). For the Bloch-type SkXs with helicity $\gamma=\pm\frac{\pi}{2}$, the strength of magnetostatic fields show a symmetric distribution over the crystal plane, and they are exactly the same, \emph{i.e.} $\bm{\mathcal{B}}_{\gamma=\frac{\pi}{2}}(\mathbf{r})=\bm{\mathcal{B}}_{\gamma=-\frac{\pi}{2}}(\mathbf{r})$, which implies that the helicity plays no role here. Moreover, the symmetry relations $\mathcal{B}_{\gamma=\pm\frac{\pi}{2},x/y}(x,y,z)=-\mathcal{B}_{\gamma=\pm\frac{\pi}{2},x/y}(x,y,-z)$ and $\mathcal{B}_{\gamma=\pm\frac{\pi}{2},z}(x,y,z)=\mathcal{B}_{\gamma=\pm\frac{\pi}{2},z}(x,y,-z)$ also exist for the components of Bloch-type SkXs, as shown in Fig.~\ref{Fig1}(c)(d). 

The magnetic SkXs can be analytically described as the hybridized triple-$\mathbf{Q}$ state, namely, the superposition of three helical states with the same pitch length and chirality on the uniform ferromagnetic magnetization $\mathbf{m}_{0}$ align along the $\hat{\mathbf{z}}$ direction 
\cite{Science2009,NatNano2013},
\begin{eqnarray}
\mathbf{m}(\mathbf{r})=\mathbf{m}_{0}\delta(z)+\mathcal{A}\sum_{i=1}^{3}[\hat{\mathbf{e}}_{z}\cos(\mathbf{Q}_{i}\cdot\mathbf{r})+\hat{\mathbf{e}}_{i}\sin(\mathbf{Q}_{i}\cdot\mathbf{r})]\delta(z).\nonumber\\\label{TripleQ}
\end{eqnarray}
Here, $\mathcal{A}$ denotes the magnetization of a single helical state; the three wavevectors $\mathbf{Q}_{i=1,2,3}$ form an angle of $2\pi/3$ with each other in the crystal plane and satisfy the relation $\sum\limits_{i=1}^{3}\mathbf{Q}_{i}=0$; $\hat{\mathbf{e}}_{z}$ is the unit vector normal to the crystal plane as defined above; $\hat{\mathbf{e}}_{i}$ are determined by the helicity $\gamma$, where $\hat{\mathbf{e}}_{i}=-\cos\gamma\hat{\mathbf{Q}}_{i}$ for N\'eel-type SkXs and $\hat{\mathbf{e}}_{i}=\sin\gamma\hat{\mathbf{e}}_{z}\times\hat{\mathbf{Q}}_{i}$ for Bloch-type SkXs.
\begin{figure*}
\includegraphics[scale=0.6]{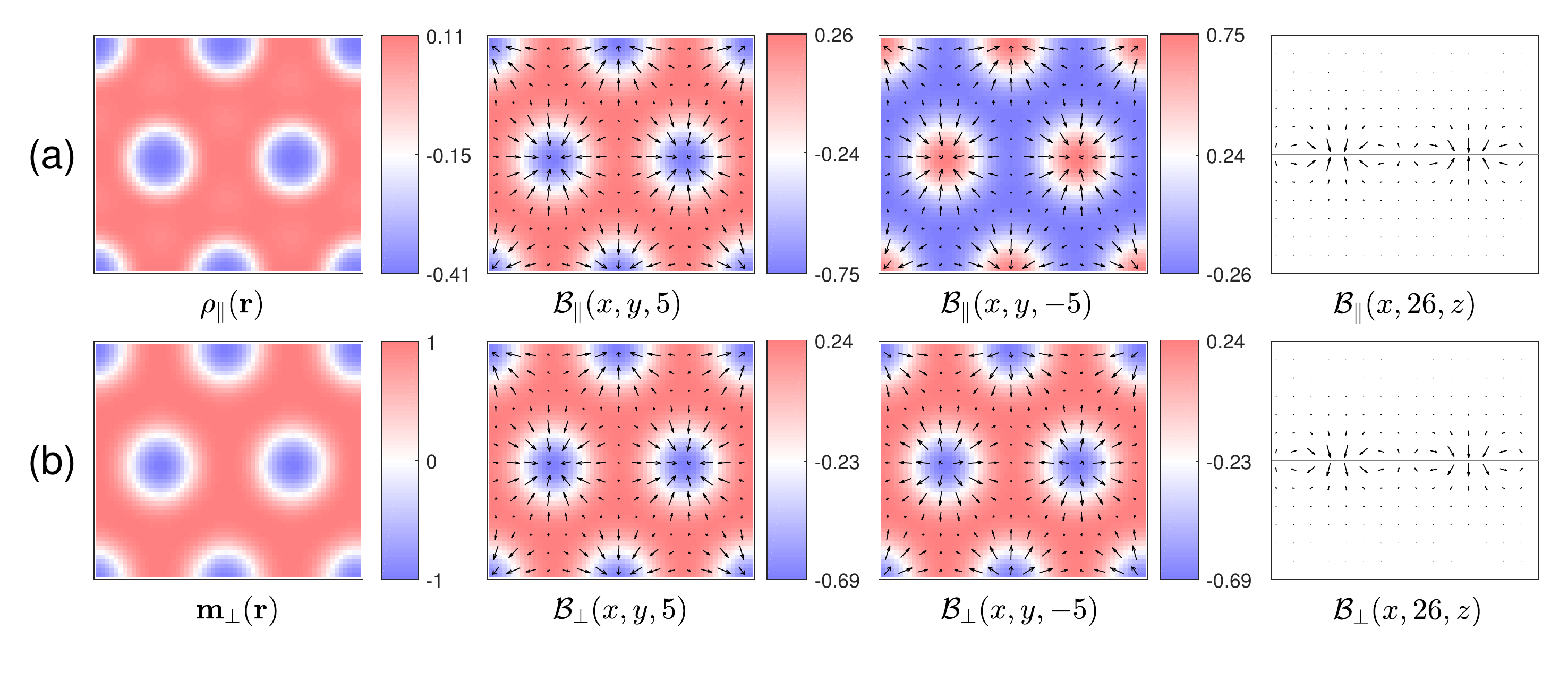}
\caption{(Color online) (a) The effective magnetic charge density of planar magnetization $\rho_{\parallel}(\mathbf{r})$ and (b) the perpendicular magnetization configuration $\mathbf{m}_{\perp}(\mathbf{r})$, and their associated magnetostatic field distributions $\bm{\mathcal{B}}_{\parallel}(\mathbf{r})$ and $\bm{\mathcal{B}}_{\perp}(\mathbf{r})$ at the $x$-$y$ planes at $z=\pm 5$ and the $x$-$z$ plane at $y=26$. The lattice constant is set as $1$ here.}\label{Fig2}
\end{figure*}
Eq.~(\ref{TripleQ}) implies that the magnetostatic field can be decomposed into two parts $\bm{\mathcal{B}}_{\perp}(\mathbf{r})$ and $\bm{\mathcal{B}}_{\parallel}(\mathbf{r})$, which are generated by the perpendicular magnetization component $\mathbf{m}_{\perp}(\mathbf{r})$ and the planar magnetization component $\mathbf{m}_{\parallel}(\mathbf{r})$ respectively. Since only the planar magnetization component is related to the helicity $\gamma$, $\bm{\mathcal{B}}_{\perp}(\mathbf{r})$ of the four types of SkXs in Fig.~\ref{Fig1} should be the same, and $\bm{\mathcal{B}}_{\parallel}(\mathbf{r})$ will be the characteristic quality to distinguish their helicities.

It is informative to understand the magnetostatic field generated by $\mathbf{m}(\mathbf{r})$ from the viewpoint of magnetic scalar potential $\Phi(\mathbf{r})$, which is defined as $\bm{\mathcal{B}}(\mathbf{r})=-\nabla\Phi(\mathbf{r})$ and is given by Poisson's equation $\nabla^{2}\Phi(\mathbf{r})=-\rho_{m}(\mathbf{r})$\cite{Jackson}. Here, $\rho_{m}(\mathbf{r})=-\nabla\cdot\mathbf{m}(\mathbf{r})$ is the effective `` magnetic charge" of the SkXs, and the vacuum permeability $\mu_{0}$ is temporally neglected for simplicity. For the planar magnetization component $\mathbf{m}_{\parallel}(\mathbf{r})$, one has
\begin{eqnarray}
\rho_{m,\parallel}(\mathbf{r})=\mathcal{A}Q\cos\gamma\sum\limits_{i=1}^{3}\cos(\mathbf{Q}_{i}\cdot\mathbf{r})\delta(z).\label{rhopar}
\end{eqnarray}
For Bloch-type SkXs with $\gamma=\pm\frac{\pi}{2}$, the magnetic charge $\rho_{m,\parallel}$ will vanish, thus the generated magnetostatic field will be soly determined by $\mathbf{m}_{\perp}(\mathbf{r})$ and is independent on the helicity. For N\'eel-type SkXs, the polarity of magnetic charge $\rho_{m,\parallel}$ will be dependent on the helicity. Therefore, the magnetostatic field of SkXs can be classified into three types according to the helicity $\gamma=0,\pi,\pm\frac{\pi}{2}$ respectively.

By solving the Poisson's equation, the magnetic scalar potential $\Phi_{\parallel}(\mathbf{r})$ generated by the planar magnetization $\mathbf{m}_{\parallel}(\mathbf{r})$ is obtained as
\begin{eqnarray}
\Phi_{\parallel}(\mathbf{r})=\cos\gamma\frac{\mathcal{A}}{2}e^{-Q|z|}\sum\limits_{i=1}^{3}\cos(\mathbf{Q}_{i}\cdot\mathbf{r}),\label{potpar}
\end{eqnarray}
then the corresponding magnetostatic field $\bm{\mathcal{B}}_{\parallel}(\mathbf{r})$ is
\begin{eqnarray}
\bm{\mathcal{B}}_{\parallel}(\mathbf{r})&=&\cos\gamma\frac{\mathcal{A}Q}{2}e^{-Q|z|}\sum_{i=1}^{3}\hat{\mathbf{n}}(\mathbf{Q}_{i},\mathbf{r}),\label{mfpar}
\end{eqnarray}
where the unit vector $\hat{\mathbf{n}}(\mathbf{Q}_{i},\mathbf{r})$ is defined as
\begin{eqnarray}
\hat{\mathbf{n}}(\mathbf{Q}_{i},\mathbf{r})=\hat{\mathbf{Q}}_{i}\sin(\mathbf{Q}_{i}\cdot\mathbf{r})+\hat{\mathbf{e}}_{z}\text{sgn}(z)\cos(\mathbf{Q}_{i}\cdot\mathbf{r}).\label{nvector}
\end{eqnarray} 
Similarly, the magnetic charge $\rho_{m,\perp}(\mathbf{r})$ for the perpendicular magnetization $\mathbf{m}_{\perp}(\mathbf{r})$ is
\begin{eqnarray}
\rho_{m,\perp}(\mathbf{r})=-(m_{0}+\mathcal{A}\sum_{i=1}^3\cos(\mathbf{Q}_{i}\cdot\mathbf{r}))\delta'(z),\label{rhoper}
\end{eqnarray} 
then the magnetic scalar potential $\Phi_{\perp}(\mathbf{r})$ and the corresponding magnetostatic field $\bm{\mathcal{B}}_{\perp}(\mathbf{r})$ will be
\begin{eqnarray}
\Phi_{\perp}(\mathbf{r})&=&\text{sgn}(z)\frac{\mathcal{A}}{2}e^{-Q|z|}\sum\limits_{i=1}^{3}\cos(\mathbf{Q}_{i}\cdot\mathbf{r}),\label{potper}\\
\bm{\mathcal{B}}_{\perp}(\mathbf{r})&=&\text{sgn}(z)\frac{\mathcal{A}Q}{2}e^{-Q|z|}\sum\limits_{i=1}^{3}\hat{\mathbf{n}}(\mathbf{Q}_{i},\mathbf{r}).\label{mfper}
\end{eqnarray}
Fig.~\ref{Fig2} shows the distributions of the effective magnetic charge density $\rho_{\parallel}(\mathbf{r})$, the perpendicular magnetization configuration $\mathbf{m}_{\perp}(\mathbf{r})$, and their associated magnetostatic fields $\bm{\mathcal{B}}_{\parallel}(\mathbf{r})$ and $\bm{\mathcal{B}}_{\perp}(\mathbf{r})$ obtained from the numerical simulations, which verify the theoretical analysis above.

The features of the magnetostatic fields shown in Fig.~\ref{Fig1} can now be well understood with Eq.~(\ref{mfpar}) and (\ref{mfper}). First, each component of the fields has the same modulation period in the $x$-$y$ plane as the underlying SkXs, and will decay exponentially with characteristic length $1/Q$ away from the crystal plane. Second, $\Phi_{\parallel}(\mathbf{r})$ and $\Phi_{\perp}(\mathbf{r})$ can be regarded as the contributions from ``inner" and ``outer" magnetic charge density, which are even and odd function of $z$ respectively, and their summation will vanish at the down(upper) half-plane for N\'eel-type SkXs with helicity $\gamma=0$ ($\gamma=\pi$); for Bloch-type SkXs ($\gamma=\pm\frac{\pi}{2}$), $\Phi_{\parallel}(\mathbf{r})$ and $\bm{\mathcal{B}}_{\parallel}(\mathbf{r})$ will vanish, and the magnetostatic fields will be the same no matter what the helicities are. In fact, the magnetization configurations of N\'eel-type SkXs form the so-called ``Halbach arrays" at nanoscale,\cite{Mallinson,Halbach,arX2017-2} which have the feature of ``one-sided flux".\cite{Mallinson} Finally, the symmetry relations of $\bm{\mathcal{B}}_{\gamma}(\mathbf{r})$ revealed in Fig.~\ref{Fig1} can be easily verified with the expressions of Eq.~(\ref{mfpar}) and (\ref{mfper}). 
  
An alternative viewpoint to understand the magnetostatic field is based on the magnetic vector potential $\mathbf{A}(\mathbf{r})$ generated by the``magnetic current density" $\mathbf{J}_{m}(\mathbf{r})=\nabla\times\mathbf{m}(\mathbf{r})$\cite{Jackson}, which is calculated to be
\begin{eqnarray}
\mathbf{J}_{m}(\mathbf{r})=\mathcal{A}Q\sum_{i=1}^{3}(\mathbf{w}(\mathbf{Q}_{i},\mathbf{r};\gamma)\delta(z)
+(\hat{\mathbf{e}}_{z}\times\hat{\mathbf{e}}_{i})\sin(\mathbf{Q}_{i}\cdot\mathbf{r})\delta'(z)),\nonumber\\\label{Jm}
\end{eqnarray}
where the vector 
$\mathbf{w}(\mathbf{Q}_{i},\mathbf{r})$ is defined as 
\begin{eqnarray}
\mathbf{w}(\mathbf{Q}_{i},\mathbf{r})=(\hat{\mathbf{e}}_{z}\times\hat{\mathbf{Q}}_{i})\sin(\mathbf{Q}_{i}\cdot\mathbf{r})+\hat{\mathbf{e}}_{z}\sin\gamma\cos(\mathbf{Q}_{i}\cdot\mathbf{r}).\nonumber\\\label{wvec}
\end{eqnarray}
In the Coulomb gauge ($\nabla\cdot\mathbf{A}=0$), the magnetic vector potential $\mathbf{A}(\mathbf{r})$ satisfies Poisson's equation $\nabla^{2}\mathbf{A}(\mathbf{r})=-\mathbf{J}_{m}(\mathbf{r})$, which results in
\begin{eqnarray}
\mathbf{A}(\mathbf{r})=\frac{\mathcal{A}}{2}e^{-Q|z|}\sum_{i=1}^{3}
(\mathbf{w}(\mathbf{Q}_{i},\mathbf{r})
-\text{sgn}(z)(\hat{\mathbf{e}}_{z}\times\hat{\mathbf{e}}_{i})\sin(\mathbf{Q}_{i}\cdot\mathbf{r}).\nonumber\\\label{VecPot}
\end{eqnarray}
Therefore, the current density $\mathbf{J}_{m}(\mathbf{r})$ and magnetic vector potential $\mathbf{A}(\mathbf{r})$ can also be decomposed into ``inner" and ``outer" contributions, which are even and odd functions of $z$ respectively. For N\'eel-type SkXs ($\gamma=0,\pi$), there is no $z$-component in $\mathbf{J}_{m}(\mathbf{r})$ and $\mathbf{A}(\mathbf{r})$, and $\mathbf{A}(\mathbf{r})$ will vanish at the down half-plane when $\gamma=0$ or upper half-plane when $\gamma=\pi$, considering that $\hat{\mathbf{e}}_{i}=-\cos\gamma\hat{\mathbf{Q}}_{i}$. For Bloch-type SkXs ($\gamma=\pm\frac{\pi}{2}$), the second term in (\ref{VecPot}) is an irrotational vector field, which suggests that the ``outer'' current density has no contribution to the magnetic field in this case. 

With the magnetic vector potential $\mathbf{A}(\mathbf{r})$ in Eq.~(\ref{VecPot}), the magnetic field $\mathbf{B}(\mathbf{r})$ is straightforwardly obtained as
\begin{eqnarray}
\mathbf{B}(\mathbf{r})&=&\frac{\mathcal{A}Q}{2}e^{-Q|z|}
\sum_{i=1}^{3}(\text{sgn}(z)\hat{\mathbf{Q}}_{i}-\hat{\mathbf{e}}_{i})\sin(\mathbf{Q}_{i}\cdot\mathbf{r})\nonumber\\
&+&(1+\text{sgn}(z)\cos\gamma)\frac{\mathcal{A}Q}{2}e^{-Q|z|}\sum_{i=1}^{3}\hat{\mathbf{e}}_{z}\cos(\mathbf{Q}_{i}\cdot\mathbf{r}).\nonumber\\
&+&\sin\gamma\frac{\mathcal{A}Q}{2}e^{-Q|z|}
\sum_{i=1}^{3}(\hat{\mathbf{e}}_{z}\times\hat{\mathbf{Q}}_{i})\sin(\mathbf{Q}_{i}\cdot\mathbf{r}).\label{Btot}
\end{eqnarray}
When $\gamma=0,\pi$, Eq.~(\ref{Btot}) will reduce to the magnetic field of N\'eel-type SkXs, \emph{i.e.} the summation of $\bm{\mathcal{B}}_{\parallel}(\mathbf{r})$ and $\bm{\mathcal{B}}_{\perp}(\mathbf{r})$; when $\gamma=\pm\frac{\pi}{2}$, Eq.~(\ref{Btot}) will reduce to the magnetic field of Bloch-type SkXs, \emph{i.e.} $\bm{\mathcal{B}}_{\perp}(\mathbf{r})$. Therefore, the results obtained from the ``magnetization current" picture are consistent with the ``magnetic charge" picture.

We now discuss the magnetostatic fields generated by stacking two chiral ferromagnetic layers, which provide us more flexibility to construct magnetic field at nanoscale. Considering that two layers with Bloch-type SkXs are located at the planes $z_{\pm}=\pm d/2$, and their magnetization configurations are $\mathbf{m}_{\pm}(\mathbf{r})$, there can be two types of magnetostatic fields between the two layers depending on the relative direction of $\hat{\mathbf{e}}_{z,\pm}$. For the parallel case with $\hat{\mathbf{e}}_{z,\pm}=\hat{\mathbf{e}}_{z}$, the magnetostatic field $\bm{\mathcal{B}}^{P}(\mathbf{r})$ is    
\begin{eqnarray}
\bm{\mathcal{B}}^{P}(\mathbf{r})&=&-\mathcal{A}Qe^{-Qd}\sum_{i=1}^{3}[\sin(\mathbf{Q}_{i}\cdot\mathbf{r})\sinh(Qz)\hat{\mathbf{Q}}_{i}\nonumber\\
&&+\cos(\mathbf{Q}_{i}\cdot\mathbf{r})\sinh(Qz)\hat{\mathbf{e}}_{z}].\label{BFP}
\end{eqnarray}
While for the antiparallel case with ${\mathbf{e}}_{z,+}=-\hat{\mathbf{e}}_{z,-}=\hat{\mathbf{e}}_{z}$, the magnetostatic field $\bm{\mathcal{B}}^{AP}(\mathbf{r})$ will be     
\begin{eqnarray}
\bm{\mathcal{B}}^{AP}(\mathbf{r})&=&-\mathcal{A}Qe^{-Qd}\sum_{i=1}^{3}[\sin(\mathbf{Q}_{i}\cdot\mathbf{r})\sinh(Qz)\hat{\mathbf{Q}}_{i}\nonumber\\
&&+\cos(\mathbf{Q}_{i}\cdot\mathbf{r})\cosh(Qz)\hat{\mathbf{e}}_{z}].\label{BFAP}
\end{eqnarray}
Eq.~(\ref{BFP}) and (\ref{BFAP}) suggest that two layers of SkXs can generate magnetostatic fields periodically modulated in the $x$-$y$ plane, and the field magnitudes depend exponentially on the layer distance $d$. With the approximate relations $e^{-Qd}\approx1$, $\sinh(Qz)\approx Qz$ and $\cosh(Qz)\approx1$ when $d\ll 1/Q$, $\bm{\mathcal{B}}^{P}(\mathbf{r})$ will be proportional to $z$ and thus has a constant gradient along the $\hat{\mathbf{e}}_{z}$ direction, while the $z$-component of $\bm{\mathcal{B}}^{AP}(\mathbf{r})$ will be much stronger than the planar component and is near constant along the $\hat{\mathbf{e}}_{z}$ direction. Besides, the magnetostatic fields can be further manipulated by translating or rotating the SkXs, which then give more types of field distributions. Considering that the magnetostatic fields of other magnetised microstructures have been successfully applied to trap and manipulate ultracold atoms in the past,\cite{JPD1999,RMP2007} we expect that SkXs would also play an unique role in atom optics.  

Finally, we estimate the amplitudes of the magnetostatic field and field gradient, which are $\mathcal{B}\sim\mu_{0}\mathcal{A}Qe^{-Qd}$ and $\nabla_{\parallel}\mathcal{B}\sim\mu_{0}\mathcal{A}Q^{2}e^{-Qd}$ respectively. Here, we retrieve the vacuum permeability $\mu_{0}=4\pi\times 10^{-7}$~T$\cdot$m/A. Assuming a ferromagnetic film with the magnetization $M_{s}\sim 1000$~kA/m, the thickness $t\sim 5$~nm, the period of SkXs $\lambda\sim 50$~nm, the layer distance $d=50$~nm and utilizing the relations $Q=2\pi/\lambda,\mathcal{A}=M_{s}t$, one gets $\mathcal{B}\sim 1.5$~mT and $\nabla_{\parallel}\mathcal{B}\sim 1.8\times10^{3}$~T/cm. By decreasing the layer distance $d$, the amplitudes $\mathcal{B}$ and $\nabla_{\parallel}\mathcal{B}$ can be further increased exponentially. Therefore, the magnetostatic fields generated by the SkXs are strong enough to trap and manipulate neutral atoms\cite{JPD1999,RMP2007}. 

In conclusion, we have revealed the features of magnetostatic fields generated by magnetic skyrmion crystals. The field generated by N\'eel-type SkX distributes only on one side of the crystal plane determined by its helicity, while the field of Bloch-type SkX distributes on both sides of the crystal plane and is irrelevant to the helicity. We have also investigated the magnetostatic field constructed by stacking two chiral ferromagnetic layers with SkXs. The results here will not only deepen our understanding of the magnetostatic characteristics of SkXs, which are important to observe SkXs with field sensing techniques and design skyrmion-based electronics devices, but also provide the possibility to trap and manipulate neutral atoms with magnetic moments at nanoscale by controlling these topological magnetic textures.  

This work is supported by NSFC Project No. 61674083 and No. 11604162.

\end{document}